\documentclass[twocolumn,aps,prl,fleqn,showpacs]{revtex4}

\usepackage{graphicx}
\usepackage{bm}
\usepackage{amsmath}
\usepackage[dvipdfm,bookmarks=true,bookmarksnumbered=true,bookmarkstype=toc]{hyperref} 

\begin{document}
\newcommand{\Area}{A}
\newcommand{\Av}{{\bm A}}
\newcommand{\av}{{\bm a}}
\newcommand{\Bv}{{\bm B}}
\newcommand{\DOS}{{\nu}}
\newcommand{\dx}{{d^3 x}}
\newcommand{\eF}{{\epsilon_F}}
\newcommand{\etaz}{{\eta^0}}
\newcommand{\etazk}{\eta^0_k}
\newcommand{\etazmk}{\eta^0_{-k}}
\newcommand{\Ev}{{\bm E}}
\newcommand{\js}{{j_{\rm s}}}
\newcommand{\jsc}{{j_{\rm s}^{\rm cr}}}
\newcommand{\jsca}{{j_{\rm s}^{\rm cr (1)}}}
\newcommand{\jscb}{{j_{\rm s}^{\rm cr (2)}}}
\newcommand{\jc}{{j^{\rm cr}}}
\newcommand{\kF}{{k_F}}
\newcommand{\kp}{{k_\perp}}
\newcommand{\Kp}{{K_\perp}}
\newcommand{\kB}{{k_B}}
\newcommand{\kBT}{{k_B T}}
\newcommand{\kv}{{\bm k}}
\newcommand{\Mv}{{\bm M}}
\newcommand{\nv}{{\bm n}}
\newcommand{\pv}{{\bm p}}
\newcommand{\Ne}{{N_{\rm e}}}
\newcommand{\Omegazk}{{\Omega^{(0)}_{\kv}}}
\newcommand{\Omegaok}{{\Omega^{(1)}_{\kv}}}
\newcommand{\omegazk}{{\omega^{(0)}_{\kv}}}
\newcommand{\Omegak}{{\Omega_{\kv}}}
\newcommand{\omegaell}{{\omega_{\ell}}}
\newcommand{\phiz}{{\phi_0}}
\newcommand{\phizc}{\bar{\phi_0}}
\newcommand{\qv}{{\bm q}}
\newcommand{\rv}{{\bm r}}
\newcommand{\rhow}{{\rho_{\rm w}}}
\newcommand{\sigmav}{{\bm \sigma}}
\newcommand{\Sv}{{\bm S}}
\newcommand{\Smh}{{\left(S-\frac{1}{2}\right)}}
\newcommand{\Tv}{{\bm T}}
\newcommand{\torq}{{\tau}}
\newcommand{\Tz}{{\tau}}
\newcommand{\tileta}{{\tilde\eta}}
\newcommand{\tilF}{\tilde{F}}
\newcommand{\tilFc}{\tilde{F}_{\rm c}}
\newcommand{\tilK}{{\tilde{K}}}
\newcommand{\tilKp}{{\tilde{K}_\perp}}
\newcommand{\tiln}{{\tilde{n}}}
\newcommand{\tilTz}{v_{\torq}}
\newcommand{\tilv}{{\tilde{v}}}
\newcommand{\thetaz}{{\theta_0}}
\newcommand{\Vz}{{V_0}}
\newcommand{\xv}{{\bm x}}
\newcommand{\xiz}{\xi_0}
\newcommand{\Vv}{{\bm V}}
\newcommand{\vv}{{\bm v}}
\newcommand{\vc}{v_{\rm c}}
\newcommand{\vs}{v_{\rm s}}
\newcommand{\vck}{v_{\rm c}^{K}}
\newcommand{\vci}{v_{\rm ins}}
\newcommand{\tilS}{\tilde{S}}
\newcommand{\beqa}{\begin{eqnarray}}
\newcommand{\eeqa}{\end{eqnarray}}
\newcommand{\beq}{\begin{equation}}
\newcommand{\eeq}{\end{equation}}
\newcommand{\la}{\langle}
\newcommand{\ra}{\rangle}

\title{Current-induced Vortex Motion by Spin-Transfer Torque}
\author{Junya Shibata}
\email{jshibata@riken.jp}
\affiliation{$^{1}$RIKEN-FRS, 2-1 Hirosawa, Wako, Saitama 351-0198, Japan\\
$^{2}$CREST, JST, 4-1-8 Honcho Kawaguchi, Saitama, Japan}

\author{Yoshinobu Nakatani}
\affiliation{University of Electro-communications, Chofu, 182-8585, Tokyo, Japan}

\author{Gen Tatara}
\affiliation{Department of Physics, Tokyo Metropolitan University,  Hachioji, Tokyo, 
192-0397, Japan\\
PRESTO, JST, 4-1-8 Honcho Kawaguchi, Saitama, Japan
}
\author{Hiroshi Kohno}
\affiliation{
Graduate School of Engineering Science, Osaka University,
Toyonaka, Osaka 560-8531, Japan}

\author{Yoshichika Otani$^{1,2}$}
\affiliation{Institute for Solid State Physics, University of Tokyo, 5-1-5 Kashiwanoha, Kashiwa, Chiba 277-8581, Japan}

\date{\today}

\begin{abstract}
We investigate the dynamics of a magnetic vortex driven by spin-transfer torque due to 
spin current in the adiabatic case. 
The vortex core represented by collective coordinate 
experiences a transverse force proportional to the product of spin current and gyrovector, 
which can be interpreted as the geometric force determined by topological charges. 
We show that this force is just a reaction force of Lorentz-type force from the spin current of conduction electrons. Based on our analyses, we propose analytically and numerically 
a possible experiment to check the vortex displacement by spin current 
in the case of single magnetic nanodot. 
\end{abstract}

\maketitle

Manipulation of nanoscale magnetization by electric current is one of the most attractive subjects 
in both basic physics and technological applications. 
After the theoretical prediction \cite{{Slonczewski},{Berger96}}, 
it has been widely recognized that spin-polarized current (spin current) plays a crucial role 
in magnetization dynamics. 
The spin current exerts a torque on localized spins 
by transferring spin angular momenta of electrons 
through the exchange interaction between conduction electrons and localized spins, 
which is thus called the spin-transfer torque.  
The key understanding of the effect is that the spin current favors magnetic configurations with spatial gradient, or more precisely, with finite Berry-phase curvature along the current. 
Such spatial gradient by spin current indeed gives rise to the motion of domain wall 
\cite{{Berger92},{TK04}}, spin-wave instability in a uniform ferromagnet 
\cite{{BJZ98},{Macdonald04},{Zhang04}} and domain nucleation \cite{STK05}.  

Recent experiments \cite{{Yamaguchi04},{Klaui05}} and numerical simulation \cite{{TNMS05}} 
for current-induced domain wall motion 
have shown that there is a vortex-like configuration in magnetic nanowire. 
Also, magnetic vortices in nanodots have drawn much attention since 
the MFM observation of a vortex core \cite{SOHSO00}. 
However, an effective force on the vortex and its dynamics by spin-transfer torque 
due to the spin current have not been clarified.

In this Letter, we present a microscopic theory of vortex dynamics 
in the presence of spin current by using the collective coordinate method. 
In the adiabatic approximation, we derive an effective force exerted 
on the vortex core due to spin-transfer torque. 
It is shown that the vortex core experiences a transeverse force, 
which compensates the Magnus-type force derived from the so-called Berry's phase term. 
This specific force is topologically invarient, 
which is characterized by topological charges. 
Unlike the case of domain wall \cite{TK04}, we show that 
there is no threshold current to induce the vortex motion 
in the absence of an external pinning. 
It is of great interest to the vortex-based devices for application.  
To verify the exisitance of current-induced transverse force on the vortex, 
we propose a possible experiment for the current-induced vortex displacement 
in a single magnetic nanodot.


We consider the Lagrangian of the localized spins in the continuum approximation. 
The spins are assumed to have easy plane taken to be $x-y$ plane, 
and are described by the Lagrangian 
\beqa
\label{Lagrangian-spins}
L_{\rm S} &=& \hbar S \int\frac{d^{3} x}{a^3}\dot{\phi}(\cos\theta-1)-H_{\rm S}, \\
H_{\rm S} &=& \frac{S^{2}}{2}\int\frac{d^{3} x}{a^3}\left\{J(\nabla{\bm n})^{2}
+ K_{\perp}\cos^{2}\theta\right\}, 
\eeqa
where ${\bm S}({\bm x},t)=S{\bm n}({\bm x},t)$ represents the localized spin vector 
with unit vector 
${\bm n}= \sin\theta\cos\phi~{\bm e}_{x}+\sin\theta\sin\phi~{\bm e}_{y}+\cos\theta~{\bm e}_{z}$ 
and the magnitude of spin $S$; ${\bm e}_{i}(i = x, y, z)$ are unit vectors of Cartesian frame.
The $J$ and $K_{\perp}$ are respectively the exchange and the 
hard-axis anisotropy constants and $a$ is the lattice constant. 
The first term of the right hand side (r.h.s) in Eq.(\ref{Lagrangian-spins}) is the so-called 
Berry-phase term, which in general determines the dynamical property of the localized spins. 


Let us denote the spin configuration of a vortex centered at the origin by a vector field 
${\bm n}_{\rm V}({\bm x})$ with unit modulus. 
As a vortex profile, we take an out-of-plane vortex; 
${\bm n}_{\rm V}({\bm x} \to {\bm 0})=p{\bm e}_{z}$, where 
$p = \pm 1$ is the polarization, which refers to the spin direction of the vortex core center 
and ${\bm n}(|{\bm x}| \gg \delta_{\rm V}) =\cos(q \varphi +C \pi/2 ){\bm e}_{x}+\sin (q \varphi +C \pi/2 ){\bm e}_{y}$, where $\delta_{V}$ is the vortex core radius, $\varphi = \tan^{-1}(y/x)$, 
$q = \pm 1,\pm2,...,$ is the vorticity, which describes the number of windings of the spin vector projected on the order-parameter space and 
$C=\pm 1$ is the chirality, which refers to the counterclockwise $(C=1)$ or the clockwise $(C=-1)$ rotational direction of the spin in the plane. 
We here introduce a collective coordinate ${\bm X}(t)=X(t){\bm e}_{x}+Y(t){\bm e}_{y}$, 
which represents vortex core center, and 
assume that a moving vortex can be written as 
${\bm n}({\bm x},t)={\bm n}_{\rm V}({\bm x}-{\bm X}(t))$ at least as a first approximation, 
that is, ignoring the spin wave excitation. 
Substituting this into Eq.(\ref{Lagrangian-spins}), 
we obtain the Lagrangian for the collective coordinate as 
\beq
\label{Lagrangian}
L_{\rm V}=\frac{1}{2}{\bm G}\cdot(\dot{{\bm X}}\times {\bm X})-U({\bm X}).
\eeq
Here ${\bm G}$ is the gyrovector defined by 
\beqa
{\bm G}={\bm e}_{z} \hbar S \int \frac{d^3 x}{a^3} {\bm n}\cdot(\partial_{x}{\bm n}\times\partial_{y}{\bm n})
=\frac{\hbar S}{a^3}2\pi L pq{\bm e}_{z}, 
\eeqa
with 
$L$ being the thickness of the system, and $U({\bm X})$ is a potential energy 
of a vortex evaluated from the Hamiltonian $H_{\rm S}$. 
The gyrovector ${\bm G}$ is topologically invariant corresponding to polarization $p$ and vorticity $q$ and the number of corvering the mapping space $D$. 
In the case of ${\bm n}_{V}({\bm x})$, 
this mapping number is $1/2$ in unit of surface area $4\pi$. 

The first term of the r.h.s in Eq.(\ref{Lagrangian}), which comes from the Berry-phase term, 
represents that $X$ and $Y$ are essentially canonically conjugate each other. 
This term provides a transeverse force $-{\bm G}\times\dot{{\bm X}}$ on the moving vortex,  
the so-called Magnus force, 
perpendicular both to the gyrovector and to the vortex velocity, whose term has been derived 
and discussed by many workers 
\cite{{Thiele73},{Volovik96},{Ao},{M-Stone96},{KY96}} in various systems. 


Let us investigate the force acting on the vortex by spin current of conduction electrons. 
The Lagrangian of the electrons are given by 
\beq
\label{electron-Lagrangian}
L_{\rm el}^{0} = 
\int d^3 x c^{\dagger}({\bm x},t)\left\{i\hbar\frac{\partial}{\partial t}
+ \frac{\hbar^{2}}{2m}\nabla^{2}\right\}c({\bm x},t)-H_{\rm sd},
\eeq
where $c(c^{\dagger})$ is anihilation (creation) operator of conduction electrons. 
The last term $H_{\rm sd}$ 
represents the exchange interaction between localized spins and conduction electrons 
given by ${H}_{\rm sd}=-\Delta\int d^3 x {\bm n}\cdot (c^{\dagger}{\bm \sigma}c)_{\bm x}$, 
where $2\Delta$ is the energy splitting, and ${\bm \sigma}$ is a Pauli-matrix vector. 
Here we perform a local gauge transformation \cite{{TF94},{TK04}} 
in electron spin space so that the quantization axis is parallel to the localized spins 
${\bm n}({\bm x},t) $ at each point of space and time;
$c({\bm x},t) = U({\bm x},t)a({\bm x},t)$, where $a({\bm x},t)$ is the two-component electron operator in the rotated frame, and $U({\bm x},t) = {\bm m}({\bm x},t)\cdot{\bm \sigma}$ is 
an SU(2) matrix with ${\bm m} = \sin (\theta/2)\cos\phi~{\bm e}_{x}+
\sin (\theta/2)\sin\phi~{\bm e}_{y}+\cos(\theta/2)~{\bm e}_{z}$. 
The Lagrangian is now given by 
\beqa
L_{\rm el} &=& \int d^3 x a^{\dagger}\bigg[
i\hbar\left(\partial_0+i A_{0}\right)\nonumber\\
&+&\frac{\hbar^2}{2m}\left(\partial_{i} + i A_{i}
\right)^2 \bigg]a + \Delta \int d^3 x a^{\dagger}\sigma_{z}a,
\eeqa
where $A_{\nu}={\bm A}_{\nu}\cdot{\bm \sigma}=-iU^{\dagger}\partial_{\nu} U~(\nu =0, x,y,z)$ is SU(2) gauge field determined by the time and spatial derivative of the localized spins. 
For slowly varying magnetic configurations, 
the electron spins can mostly follow adiabatically the localized spins. 
This is justified for the condition $(\Delta/\varepsilon_{\rm F})k_{\rm F}\lambda\gg 1$, 
where $\varepsilon_{\rm F}$ and $k_{\rm F}$ are, respectively, 
the Fermi energy and Fermi wave number of conduction electrons, 
and $\lambda$ is the characteristic length scale of the spin texture of the localized spins. 
In this adiabatic approximation, 
taking the expectation value of $L_{\rm el}$ 
for the current-carrying nonequilibrium state, 
we can obtain the following 
interaction Hamiltonian of the first-order contribution to the localized spins \cite{STK05};
\begin{equation}
\label{spin-torque-Hamiltonian}
H_{\rm ST}=\int d^3 x \frac{\hbar}{2e}{\bm j}_{\rm s}\cdot\nabla\phi\cdot(1-\cos \theta),
\end{equation}
where ${\bm j}_{\rm s}$ is the spin current density, 
which is written by using the distribution function 
$f_{{\bm k}\sigma}=\langle a^{\dagger}_{{\bm k}\sigma}a_{{\bm k}\sigma}\rangle$ 
in the rotated frame specifying the current-carrying nonequilibrium state as 
$
{\bm j}_{\rm s} 
= \frac{1}{V}\sum_{{\bm k},\sigma}\sigma \frac{\hbar{\bm k}}{m}f_{{\bm k}\sigma}.
$
As seen from Eq. (\ref{spin-torque-Hamiltonian}), 
the spin current favors a finite Berry-phase curvature along the current. 
Indeed, the Hamiltonian $H_{\rm ST}$ leads to the spin-transfer torque in the case of domain wall 
\cite{STK05}. 

From Eq.({\ref{spin-torque-Hamiltonian}), 
we can derive the effective Hamiltonian represented by the collective coordinate 
for the vortex core. 
Taking the variation $\delta{\bm n} = -\partial_{j}{\bm n}_{\rm V}({\bm x}-{\bm X})\delta X_{j}$, 
where repeated roman indices imply sum over the in-plane spatial direction $j=x,y$, 
we obtain 
\beq
\label{deviation}
\delta H_{\rm ST} = 
\frac{\hbar S}{a^3}2\pi L p q (v_{{\rm s},x}\delta Y-v_{{\rm s},y}\delta X), 
\eeq
where ${\bm v}_{\rm s} = \frac{a^{3}}{2eS}{\bm j}_{\rm s}$ 
represents the drift velocity of electron spins. 
By integrating Eq.(\ref{deviation}), we obtain 
\beq
H_{\rm ST} = {\bm G}\cdot({\bm v}_{s}\times{\bm X}). 
\eeq 
Thus, a force acting on the vortex core is given by 
\beq
\label{spin-torque-force}
{\bm F}_{\rm ST} = - \frac{\partial H_{\rm ST}}{\partial {\bm X}} = -{\bm G}\times {\bm v}_{\rm s}. \eeq
This current-induced transverse force has been previously derived by Berger 
\cite{Berger86} based on a phenomenological treatment in the case of Bloch line. 
Here we have derived this force from microscopic theory. 
It is noted that this force does not depend on the chirality $C=\pm 1$ of the vortex 
in contrast to a force produced by a magnetic field. 
Since it is hard to control the chirality, this fact would great advantage in application.  
A microscopic derivation of a general relation between force and torque 
in the Landau-Lifshitz-Gilbert (LLG) equation is presented in Ref. \cite{KTS05}. 

Before proceeding, we here briefly remark that this force can be interpreted as a reaction force 
of a Lorentz-type force from the spin current of conduction electrons. 
The interaction Hamiltonian, Eq.(\ref{spin-torque-Hamiltonian}), 
can be rewritten by using the 
the $U(1)$ gauge field,  
${\bm A}({\bm x})=\frac{\hbar}{2e}\nabla\phi({\bm x})(\cos\theta({\bm x})-1)$, 
interacting with the spin current density. 
Thus the magnetic field ${\bm B}({\bm x})$ is given by 
\begin{equation}
\label{magnetic-field}
{\bm B} = (\partial_{x}A_{y}-\partial_{y}A_{x}){\bm e}_{z}
=-\frac{\hbar}{2e}{\bm n}\cdot(\partial_{x}{\bm n}\times\partial_{y}{\bm n}){\bm e}_{z}, 
\end{equation}
which corresponds to the so-called topological field \cite{Bruno04}.
It is noted that the SU(2) field intensity, 
$F_{\mu\nu}=\partial_{\mu}A_{\nu}-\partial_{\nu}A_{\mu}-i[A_{\mu},A_{\nu}]$,  
vanishes, since the original Lagrangian $L^{0}_{\rm el}$ 
in Eq.(\ref{electron-Lagrangian}) does not include the local gauge field. 
The finite magnetic field is a consequence of 
the projection from SU(2) to the U(1)  
by taking the adiabatic approximation. 
Under this magnetic field, 
the spin current of the conduction electrons may experience 
the following Lorentz-type force as 
\beqa
{\bm F}_{\rm L} = \int d^3 x ~{\bm j}_{\rm s}\times{\bm B}
={\bm G}\times {\bm v} _{\rm s} = -{\bm F}_{\rm ST},
\eeqa
which can be interpreted as the reaction force acting on the vortex. 

Let us derive the equation of motion for the collective coordinate of vortex  
in the presence of spin current based on the Euler-Lagrange equation; 
\beqa
\label{Euler-Lagrange}
\frac{d}{dt}\frac{\partial L}{\partial\dot{{\bm X}}}-\frac{\partial L}{\partial{\bm X}}
=-\frac{\partial W}{\partial \dot{{\bm X}}}, 
\eeqa
where $L = L_{\rm V}-H_{\rm ST}$ is the total Lagrangian, 
$W$ is the so-called dissipation function given by 
\beqa
W = \alpha\frac{\hbar S}{2} \int \frac{d^3 x}{a^3}~\dot{{\bm n}}^{2}({\bm x},t) = \frac{\alpha}{2}D\dot{{\bm X}}^2,
\eeqa
with  $\alpha$ being the Gilbert damping constant.
The constant 
\beqa
D = \frac{\hbar S}{a^{3}}L \int_{D}dxdy 
\left\{(\partial_{i}\theta)^2+\sin^{2}\theta (\partial_{i}\phi)^2\right\}, 
\eeqa
generally includes a factor 
$\ln (R_{\rm V}/\delta_{\rm V})$, where $R_{\rm V}$ is the system size. 
Here we assume that system has rotational invariance along the $z$ axis. 
The concrete expression in Eq. (\ref{Euler-Lagrange}) is given by 
\beqa
\label{vortex-equation}
{\bm G}\times ({\bm v}_{\rm s} -\dot{\bm X}) = 
 -\frac{\partial U({\bm X})}{\partial {\bm X}} 
-\alpha D\dot{\bm X}. 
\eeqa
This is the equation of motion for the vortex dynamics in the presence of spin current. 
If the r.h.s in Eq.(\ref{vortex-equation}) is absent, we obtain 
$\dot{\bm X} = {\bm v}_{\rm s}$, 
where the vortex core moves along the spin current 
perpendicular to the transverse force ${\bm F}_{\rm ST}$. 
This situation can be seen from the current-induced domain wall motion \cite{{TK04},{STK05}}. 
On the other hand, the damping term $-\alpha D \dot{{\bm X}}$ acts as 
deviating from the orbital direction of the moving vortex along the current. 
Importantly, there is no intrinsic pinning in the dynamics of vortex 
unlike the case of domain wall. 
This lead to a vanishing threshold current for the vortex motion 
in the absence of an external pinning. 
This is because, in the translationally-invariant system, 
there is no pinning on $X$ and $Y$. 
This is in contrast with the case of domain wall, 
where, $\phi_{0}$ is pinned by the hard-axis magnetic anisotropy 
even in the translationally-invariant system \cite{TK04}. 
Thus vortex-based devices would have great advantage in low-current opperation. 

To verify the exisitence of current-induced force on the vortex, 
we propose the vortex displacement by spin current in 
single magnetic nanodot \cite{VGMZL05}, where an out-of-plane vortex 
with vorticity $q=1$ is stabilized. 
We assume the electric current is uniform in the nanodot, 
and flowing in the possitive x-direction, that is, ${\bm v}_{\rm s} = 
\frac{a^3}{2eS}{\bm j}_{\rm s} = v_{\rm s}{\bm e}_{x}$.
We assume the full spin polarization of the current, $P=1$, for simplicity (Fig. 1). 
\begin{figure}[bp]
\includegraphics[scale=0.4]{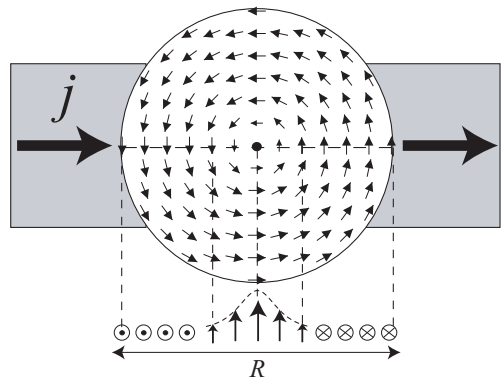}
\caption{Schematic  illustration 
of an experimental set up for current-induced vortex displacement. 
The topological charges are chosen to be $p=1$, $q=1$, and $C=1$ in the above vortex.}
\label{fig1}
\end{figure}
The potential energy $U({\bm X})$ is modeled by a harmonic one 
$U({\bm X}) = \kappa {\bm X}^{2}/2$, where $\kappa$ is a force constant. 
In Ref. \cite{Gu02},  $\kappa$ is evaluated in detail, 
which depends on the aspect ratio $g=L/R$, where $R$ is the dot radius. 
From Eq.(\ref{vortex-equation}), the equation of motion is given by 
\beqa 
(1+i\tilde{\alpha})\dot{Z}=-i\omega Z + v_{\rm s}, 
\eeqa
where $Z = X + i Y$, $\tilde{\alpha}= \alpha D/G$, and $\omega = \kappa / G$. 
For an initial condition $Z(0)= 0$, the solution is given by 
\beqa
Z(t) = i\frac{v_{s}}{\omega}\left\{
\exp\left(-\frac{i\omega t}{1+i\tilde{\alpha}} \right)-1
\right\}.
\eeqa
Thus the vortex center exhibits a spiral motion, 
whose rotational direction depends on the sign of the core polarization $p=\pm 1$ (Fig. 2(a)). 
The final displacement of the vortex core is perpendicular to the current direction 
and given by $\delta Y = - G v_{\rm s}/\kappa$, which also depends on $p$,  
where the transverse force $-{\bm G} \times {\bm v}_{\rm s}$ 
balances the restoring force $-\kappa {\bm X}$. 
This result is consistent with the recent experimental one in Ref. \cite{Klaui05}, 
where a distorted vortex-wall was shifted to the diagonal direction and stopped moving. 

\begin{figure}[bp]
\includegraphics[scale=0.4]{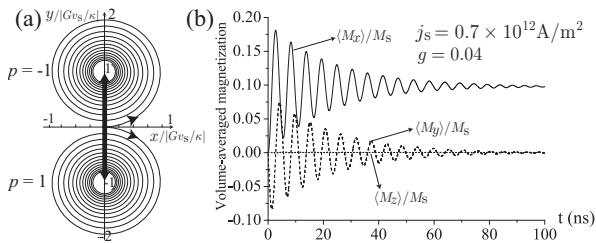}
\caption{(a)Spiral motion of vortex center under spin current obtained from the analtycal calculations. We took $\tilde{\alpha}= 0.02$.
(b)Numerical results of the time evolution of the volume-averaged magnetization components.}
\label{fig2}
\end{figure}

For comparison, the current-induced dynamics of a vortex was 
calculated by micromagnetic numerical simulations based on 
the LLG equation with spin current terms; 
\beq
\label{modified-LLG}
\frac{\partial{\bm M}}{\partial t} =-\gamma_{0}{\bm M}\times{\bm H}_{\rm eff}
+\frac{\alpha}{M_{\rm s}}
{\bm M}\times \frac{\partial{\bm M}}{\partial t}-{\bm v}_{\rm s}\cdot\nabla{\bm M},
\eeq
where ${\bm M}({\bm x},t)$ is the magnetization vector, 
$\gamma_{0}$ is the gyromagnetic constant, 
${\bm H}_{\rm eff}$ is the effective magnetic field 
including the exchange and demagnetizing field 
and $M_{\rm s}$ is the sarturation magnetization. 
The last term represents the spin-transfer torque 
\cite{{BJZ98},{Zhang04},{Macdonald04},{STK05}}. 
The sample is divided into identical cells, 
in each of which magnetization is assumed to be constant. 
The dimension of the cells is $4\times4\times h ~{\rm nm}^{3}$ 
with dot thickness $h= 10,~20,~30~{\rm nm}$. 
The dot radius is taken to be $R = 500~{\rm nm}$. 
The computational material parameters are typical for permalloy; 
$M_{\rm s} = 1.0 ~{\rm T}~(M_{\rm s}/\mu_{0} = 8.0 \times 10^{5}~{\rm A/m})$, 
the exchange stiffness constant $A=1.0 \times 10^{-11}~{\rm J/m}$ 
and $\gamma_{0} = 1.8 \times 10^{5} {\rm m}/{\rm A\cdot s}$. 
We take $\alpha=0.02$. 
The programing code is based on those of Refs. \cite{{TNMS05},{NTM03}}. 

Figure 2(b) shows the time evolution of volume-averaged magnetization, 
which exhibits the spiral motion of vortex core. 
It is noted that the rotational direction is opposite to the case of 
analytical result because of the replacemet $\hbar S / a^3 \to - M_{\rm s}/\gamma_{0}$. 
Figure 3 shows that vortex displacement as a function of the spin current density 
in various aspect ratios $g=L/R$. 
The numerical results are in good agreement with the analytical ones 
for small vortex displacement. 
It is found that the smaller $g$ is more advantageous to the vortex displacement. 
\begin{figure}[bp]
\includegraphics[scale=0.4]{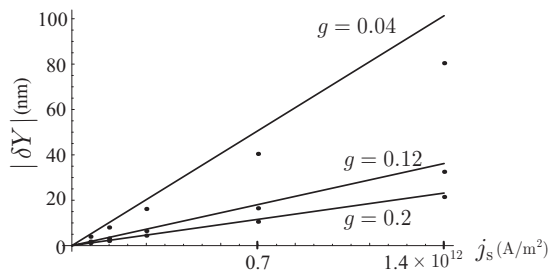}
\caption{Displacement of vortex center as a function of the spin current density $j_{\rm s}$ 
in various aspect ratios $g$. 
Solid lines and dots represent the analytical and numerical results, 
respectively.}
\label{fig3}
\end{figure}

In conclusion, 
we have clarified the transverse force on the vortex 
and its dynamics by spin-transfer torque due to the spin current 
in the adiabatic regime. 
We have proposed analytically and numerically 
a possible experiment for the vortex displacement by spin current 
in the case of single magnetic nanodot. 
Finally, this vortex displacement may affect a magnetoregistance, which 
is probable to be detected, for example, by using planar Hall effect \cite{IKO05}.

The authors would like to thank T. Ishida and T. Kimura for valuable discussion.


\begin{thebibliography}{99}

\bibitem{Slonczewski}
J. C. Slonczewski, J. Magn. Magn. Mater. {\bf 159}, L1 (1996). 

\bibitem{Berger96}
L. Berger, Phys. Rev. B {\bf 54}, 9353 (1996). 



\bibitem{Berger92}
L. Berger, J. Appl. Phys. {\bf 71}, 2721 (1992); 
E. Salhi and L. Berger, {\it ibid}. {\bf 73}, 6405 (1993).

\bibitem{TK04}
G. Tatara and H. Kohno, Phys. Rev. Lett. {\bf 92}, 086601 (2004).


\bibitem{BJZ98}
Ya. B. Bazaliy, B. A. Jones, and Shou-Cheng Zhang, 
Phys. Rev. B {\bf 57}, R3213 (1998). 


\bibitem{Macdonald04}
J. Fern\'andez-Rossier, M. Braun, A.S. N\'u\~nez and A. H. MacDonald, 
Phys. Rev. B {\bf 69}, 174412 (2004). 

\bibitem{Zhang04}
Z. Li and S. Zhang, Phys. Rev. Lett. {\bf 92}, 207203 (2004). 


\bibitem{STK05}
J. Shibata, G. Tatara and H. Kohno, Phys. Rev. Lett. {\bf 94}, 076601 (2005). 


\bibitem{Yamaguchi04}
A. Yamaguchi, T. Ono, S. Nasu, K. Miyake, K. Mibu, and T. Shinjo, 
Phys. Rev. Lett. {\bf 92}, 077205 (2004). 


\bibitem{Klaui05}
M. Kl${\rm \ddot{a}}$ui, P. O. Jubert, R. Allenspach, A. Bischof, 
J. A. C. Bland, G. Faini, U. R${\rm \ddot{u}}$diger, C. A. F. Vaz, 
L. Vila, and C. Vouille, Phys. Rev. Lett. {\bf 95}, 026601 (2005).




\bibitem{TNMS05}
A. Thiaville, Y. Nakatani, J. Miltat and Y. Suzuki, 
Europhys. Lett. {\bf 69}, 990 (2005). 

\bibitem{SOHSO00}
T. Shinjo, T. Okuno, R. Hassdorf, K. Shigeto, and T. Ono, 
Science {\bf 289}, 930 (2000). 


\bibitem{TF94}
G. Tatara and H. Fukuyama Phys. Rev. Lett. {\bf 72}, 772 (1994); 
J. Phys. Soc. Jpn. {\bf 63}, 2538 (1994). 


\bibitem{Berger86}
L. Berger, Phys. Rev. B {\bf 33}, 1572 (1986). 

\bibitem{Thiele73}
A. A. Thiele, Phys. Rev. Lett. {\bf 30}, 230 (1973). 

\bibitem{Volovik96}
G. Volovik, JETP Lett. {\bf 44}, 185 (1986). 

\bibitem{Ao}
P. Ao and D. J. Thouless, Phys. Rev. Lett. {\bf 70}, 2158 (1993). 

\bibitem{M-Stone96}
M. Stone, Phys. Rev. B{\bf53}, 16573 (1996).

\bibitem{KY96}
H. Kuratsuji and H. Yabu, 
J. Phys. A {\bf 29}, 6505 (1996);
H. Kuratsuji and H. Yabu, {\it ibid}. {\bf 31}, L61 (1998). 

\bibitem{KTS05}
H. Kohno, G. Tatara and J. Shibata, in preparation. 


\bibitem{Bruno04}
P. Bruno, V. K. Dugaev and M. Taillefumier, 
Phys. Rev. Lett. {\bf 93}, 096806 (2004). 


\bibitem{VGMZL05}
P. Vavassori, M. Grimsditch, V. Metlushko, N. Zaluzec, and B. Llic, 
Appl. Phys. Lett. {\bf 86}, 072507 (2005). 

\bibitem{Gu02}
K. Yu. Guslienko, B. A. Ivanov, V. Novosad, Y. Otani, H. Shima and K. Fukamichi, 
J. Appl. Phys. {\bf 91}, 8037 (2002). 

\bibitem{NTM03}
Y. Nakatani, A. Thiaville and J. Miltat, 
Nature materials {\bf 2}, 521 (2003).

\bibitem{IKO05}
T. Ishida, T. Kimura and Y. Otani, in preparation. 



























 








\end{thebibliography}
\end{document}